\documentclass[twocolumn,a4paper,10pt]{extarticle}

\usepackage{mathptmx}

\usepackage[compact]{titlesec}

\titleformat{\section}[block]
{\fontsize{14}{15}\bfseries}
{\thesection}
{1em}
{}
\titleformat{\subsection}[block]
{\fontsize{11}{15}\bfseries}
{\thesubsection}
{1em}
{}

\usepackage[margin=0.78in]{geometry}

\usepackage{amsmath}
\usepackage{newfloat}
\usepackage{listings}
\usepackage{balance}

\usepackage[final]{microtype}
\usepackage[sort,nocompress]{cite}

\usepackage{graphicx}

\usepackage{algorithmic}

\usepackage{booktabs,tabularx}

\usepackage[hyphens]{url}

\usepackage{subcaption}

\usepackage{mathtools}
\usepackage{centernot}

\def\name{GhostMinion}

\newcommand{\myparagraph}[1]{\vskip .4ex\noindent\textbf{#1}\hspace{.6em}}

\def\order{Strictness Order}

\def\orderh{Strictness-Order}
\def\lorder{Temporal Order}
\def\lorderh{Temporal-Order}

\usepackage{cleveref}
\usepackage{framed} 
\usepackage{amssymb,amsmath,mathtools, amsthm}
\usepackage[framemethod=tikz]{mdframed}
\mdfdefinestyle{mystyle}{
	hidealllines=true,
	leftline=true,
	innerleftmargin=10pt,
	innerrightmargin=10pt,
	innertopmargin=0pt,
}

\newmdenv[
topline=false,
bottomline=false,
skipabove=\topsep,
skipbelow=\topsep
]{siderules}

\newmdtheoremenv[style=mystyle]{frm-def}{Definition}

\crefname{frm-def}{definition}{definitions}

\newcommand{\response}[1]{#1}
\title{\name{}: A \orderh ed Cache System for Spectre Mitigation}

\author{Sam Ainsworth \\  University of Edinburgh \\  \textit{sam.ainsworth@ed.ac.uk}
}

\date{}   

\begin{document}

\maketitle

	\begin{abstract}
		
		Out-of-order speculation, a technique ubiquitous since the early 1990s, remains a fundamental security flaw. Via attacks such as Spectre and Meltdown, an attacker can trick a victim, in an otherwise entirely correct program, into leaking its secrets through the effects of misspeculated execution, in a way that is entirely invisible to the programmer's model. This has serious implications for application sandboxing and inter-process communication.
		
		Designing efficient mitigations that preserve the performance of out-of-order execution has been a challenge. The speculation-hiding techniques in the literature have been shown to not close such channels comprehensively, allowing adversaries to redesign attacks. Strong, precise guarantees are necessary, but mitigations must achieve high performance to be adopted. We present \order ing, a new constraint system that shows how we can comprehensively eliminate transient side channel attacks, while still allowing complex speculation and data forwarding between speculative instructions. We then present \name, a cache modification built using a variety of new techniques designed to provide \order\ at only 2.5\% overhead.
		
	\end{abstract}

	
	\section{Introduction}
	
	Speculative side-channel attacks such as Spectre~\cite{Kocher2018spectre} and Meltdown~\cite{Lipp2018meltdown} have severely damaged modern systems' security: even functionally correct programs can be compromised through the leakage of secrets via incorrectly predicted program behaviour. Still, the side channels that cause this leakage, for example caches, are desirable in many cases; the speed of modern microarchitectures is only possible because of structures that leak soft state. Out-of-order speculation, which allows attackers to execute complex speculative code, is mandatory for high instruction-level parallelism. 
	
	Many different mitigation techniques, with differing threat models and security properties, have been proposed.
	Often the tradeoffs are unclear, as the guarantees necessary, and provided, are typically opaque. 
	This has resulted in subtle bugs in many proposed countermeasures. Since precise models of the behaviour guaranteed by mitigations are typically unavailable, sophisticated attacks to circumvent restrictions, such as SpectreRewind~\cite{fustos2020spectrerewind} and Speculative Interference~\cite{behnia2020speculative} have appeared to fill in the gaps. 
	We now know that cleaning up the effects of speculation after it has occurred is insufficient to prevent microarchitectural speculation attacks. This means that speculation-hiding techniques in the literature~\cite{Saileshwar:2019:CUA:3352460.3358314,Ainsworth_2020,wu2020reversispec,yan2018invisispec} leak, through inability to correctly handle concurrent execution and incomplete coverage of microarchitectural structures. Examples include but are not limited to allowing structural hazards in non-pipelined  divider units to be caused by otherwise ``invisible'' speculative loads~\cite{fustos2020spectrerewind}, and can cause comprehensive breaches, and leave long-lasting traces in otherwise protected cache systems~\cite{behnia2020speculative}.
	Speculation-restricting techniques~\cite{STT-micro,NDA-micro,SDVP,CondSpec}, by comparison, prevent many executions that are completely valid, with significant performance loss for memory-intensive workloads.
	
	We present \order ing, a permissive constraint system that allows information (and side-channel) flow in any situation where speculative operations that do not commit can never leak to those that do. This is complicated by the fact those instructions may execute concurrently~\cite{behnia2020speculative,fustos2020spectrerewind}, but can be expressed as a simple property (\cref{def:order}). This allows significantly more freedom than only allowing instructions to emit timing side-effects once non-speculative~\cite{STT-micro,NDA-micro,SDVP,CondSpec}, because whether an instruction commits or not is heavily tied with the fates of other concurrently executing instructions. For example, instructions in a pipeline only commit when all previous instructions in program order do so, and so those previous instructions can freely transmit data to, and take resources from, concurrently executing future instructions. 
	
	We show how to build \order ing into microarchitecture with \name, a cache system and core modification designed to eliminate transient execution attacks including Spectre~\cite{Kocher2018spectre}, and associated derivatives including SpectreRewind~\cite{fustos2020spectrerewind} and Speculative Interference~\cite{behnia2020speculative}. Overheads \response{from the cache modification} are just $2.5\%$ on SPEC CPU2006, $0.6\%$ on SPECspeed 2017, and negligible on Parsec, \response{and we argue that fixing other \orderh\ issues within the system are unlikely to have significant overhead}. \name\ defeats transient execution attacks more comprehensively than previous speculation-hiding defences~\cite{Ainsworth_2020,Saileshwar:2019:CUA:3352460.3358314,wu2020reversispec,10.1145/3310273.3321558,gonzalez2019replicating,yan2018invisispec,DBLP:journals/corr/abs-1806-05179}, and without fully associative overprovisioned structures~\cite{yan2018invisispec,DBLP:journals/corr/abs-1806-05179}. 
	
	Our contributions are as follows:
	
	\begin{itemize}
		\item We define a timing model, \order, that if obeyed can eliminate all forms of transient execution attack, including backwards-in-time and contention attacks~\cite{Ainsworth_2020,fustos2020spectrerewind,behnia2020speculative}, while still being highly permissive to innocuous forms of information flow and out-of-order execution.
		\item We define \lorder\ as a simple overapproximation of \order.
		\item We develop \name, a cache system that uses the new techniques of TimeGuarding, Lapfrogging and Timeleaping to enforce \lorder, and thus \order, in the cache hierarchy.
		\item We implement or describe the extensions that would be necessary to achieve \lorder\ and/or \order\ for the entire system, for hypothetical attacks that violate \order\ (via contention or soft state) anywhere within a processor, and argue that the techniques demonstrated by \name\ can be used elsewhere and the additional overheads are likely to be neglgible.
		\item We evaluate \name\ in gem5, showing overheads of only 2.5\% geomean for SPEC CPU2006, 0\% for Parsec, and 0.6\% for SPECspeed 2017.
	\end{itemize}
	
	\subsection{Threat Model}
	
	We assume the existence of an attacker, trying to leak secret data through transient (misspeculated) execution. They will succeed if they can change the execution time of \textit{any} committed instruction at any point before, during or after the misspeculation, via a transient access of the secret they are trying to leak. This measurement includes but is not limited to the execution time of instructions that directly measure time, such as \texttt{rdtsc}. The attack may be via attacks such as Spectre~\cite{Kocher2018spectre}, where no explicit exceptional behaviour is triggered by the misspeculation, or attacks such as Meltdown~\cite{Lipp2018meltdown} or Foreshadow~\cite{217543} where exceptional behaviour is incorrectly temporarily ignored by the speculative execution, which can be fixed more directly. When we refer specifically to Spectre, we do so because it requires more all-encompassing strategies for defence, and without loss of generality.
	
	We do not necessarily require the attacker to successfully bring in or evict data from any observable cache past the point of misspeculation. This is because recent attacks~\cite{fustos2020spectrerewind} are able to measure secrets just from the effects on concurrently executing instructions, and such contention effects are able to be converted into long-lasting state changes~\cite{behnia2020speculative}. 
	
	We assume the attacker is able to run arbitrary code, but cannot load secret data directly; e.g., they are within a sandbox either in userspace or the kernel, sharing an address space with the secret but with runtime checks preventing access. The attacker could also reside in another process~\cite{Ainsworth_2020}, or on another system entirely~\cite{NetSpectre}.
	
	
	\section{Background}
	
	\subsection{Spectre}
	
	Spectre~\cite{Kocher2018spectre} uses transient misspeculation of a processor, such as from a mistrained branch predictor or branch target buffer, to accesse information that the correct execution of the program would not. This can be used by an attacker to trick a victim into loading secret data, for example by ignoring safety checks on inputs or within sandboxes. Though the execution is rolled back, it may leave traces within soft-state structures such as caches, which can be measured using timing side channels to reveal the secret.
	
	In the original Spectre proof-of-concept~\cite{Kocher2018spectre}, the attacker collects secret data by using it to then trigger other transient loads into the cache based on using the secret value as an index. This is followed by measuring which cache lines have been brought into the cache once correct execution restarts, in an evict-and-time attack, or evicted, via a prime-and-probe attack. This gave rise to defences that cleared up misspeculation effects after discovery~\cite{Ainsworth_2020,gonzalez2019replicating,10.1145/3310273.3321558,Saileshwar:2019:CUA:3352460.3358314,wu2020reversispec}.
	
	\subsection{Backwards-in-Time Attacks}
	
	Post-misspeculation rollback is insufficient against attacks that rely on changing the timing of committed instructions \textit{before} the misspeculation, such as SpectreRewind~\cite{fustos2020spectrerewind}. Out-of-order processors execute many instructions in parallel. Though the frontend fetches instructions in sequence, and the backend retires them in sequence, in the middle, operations that do not commit (and are misspeculated) can occur before those that do, which are earlier in program order. SpectreRewind~\cite{fustos2020spectrerewind} uses misspeculation to change the timing of these logically earlier (but executed later) instructions, by causing structural-hazard contention in non-pipelined divider units. Speculative Interference~\cite{behnia2020speculative} takes this one step further: it notes that these timing effects on logically earlier instructions can actually change the order in which instructions execute, causing different state to be stored in the cache system, and thus leaving a long-lasting trace of behaviour that can later be picked up by an attacker. The effect is not limited to non-pipelined functional units; equivalent contention can be caused by transient instructions bringing data into, or evicting data from, the cache for earlier concurrently executing instructions to use~\cite{Ainsworth_2020}. Clearing state changes on the discovery of misspeculation~\cite{Ainsworth_2020,gonzalez2019replicating,10.1145/3310273.3321558,Saileshwar:2019:CUA:3352460.3358314,wu2020reversispec} is insufficient: side channels can travel \textit{backwards in time} to cause equivalent damage.

	\section{Strictness Ordering}
	\label{sec:order}

	Speculative-execution attacks cause measurable behaviour that can be exploited to retrieve secrets. This is achieved by causing transient instructions, which would never be executed without the presence of misspeculation, to affect the timing of other instructions that are otherwise correct. To complicate this, instructions within a pipeline are not cleanly split up in to ``transient'' and ``non-speculative'', where the latter can transmit to any instruction and the former cannot: instead, a pipeline is filled with instructions from progressive levels of speculation. During execution it is typically unknown how far down the path of instructions we will commit before a transient misspeculation, and instruction flush, will occur.
	
	Still, we must not be overly restrictive, in capturing innocuous transmission between instructions that cannot leak transient execution to committed state.
	To directly capture the threat of instructions that do not commit affecting the timing of those that do, we define a new timing influence model that must be obeyed to prevent speculative timing-side-channel attacks (\cref{def:order}).
	\vspace{5pt}
	\begin{frm-def}[\order ing]
		
		Let \emph{x} and \emph{y} denote executed instructions. Under \order ing, \response{\emph{y} can \emph{strictly observe} \emph{{x}}} ($x \xRightarrow{\text{S}} y$), meaning \emph{x} can impact the execution time  (cycle of any pipeline stage) of \emph{y}, iff 	
		\[ commit(y) \rightarrow commit (x)\]
		where $commit (i)$ is true iff instruction $i$ is guaranteed to reach the end of the CPU pipeline without being squashed, or already has, and $\rightarrow$ is logical implication.
		\label{def:order}	
	\end{frm-def}
	
	In other words, for $y$ to observe any timing change from the execution of $x$, $y$ must be an instruction that is always squashed whenever $x$ is squashed, or $x$ must be an instruction that will  not be squashed (or has already committed). Intuitively, this can be viewed as preventing timing side-channel effects from flowing \textit{backwards-in-time} from more speculative to less speculative instructions.
	
	\begin{siderules}
		\response{To prove this is sufficient to eliminate transient side-channel attacks, suppose $x$ is an instruction that is part of a transient-execution gadget: it can observe secret data $D$ that is inaccessible under correct execution, and thus never commits, so $\neg commit(x)$.}
		\response{Since $x$ does not commit, neither will any instructions that depend on it, and so $D$ may only be recovered via a (timing) side channel. If we assume the attack succeeds, some instruction $y$ exists, such that $commit(y)$ holds, that has its timing impacted by the execution of $x$ to measure the channel\footnote{This may be a timer such as \texttt{rdtsc}, or be based on the commit time of an instruction with state, such as an incrementing counter~\cite{timers}.}. 
			For that to have occured, $x \xRightarrow{\text{S}} y$  holds by definition, and so $commit(y) \rightarrow commit (x)$. Since $commit(y)$ is } \response{true, $commit(x)$ must also hold. However, since x is transient, $\neg commit(x)$ also holds, and we have a contradiction: no transient $x$ that can transmit data to a committed instruction via timing side channel exists.}
	\end{siderules}
	
	
	There are a number of properties that follow:

	\myparagraph{Every instruction that has already committed, or is guaranteed to commit, can transmit to any other.} Side-channel flow will always eventually be able to occur for valid instructions executing in program-order, as a fallback for more advanced reordering.
	\myparagraph{Within a thread and pipeline, for any two instructions $a$ and $b$, either $a \xRightarrow{\text{S}} b$, $b \xRightarrow{\text{S}} a$, or both.} Assume that, as is typical, an out-of-order processor issues instructions (speculatively) in order at the front-end, and that misspeculation is handled by restarting the pipeline at the last known correct instruction in program order. It follows that there is a total order on instructions that commit: for an instruction to commit, all earlier instructions issued in program order must also do so, thus at least one of $commit(a) \rightarrow commit(b)$ or $commit(b) \rightarrow commit(a)$ holds.  This allows us to be more lenient than waiting for instructions to become non-speculative, because it always allows older instructions to immediately transmit to newer instructions in program order, even while speculative. 
	\myparagraph{The relation is transitive.} Imagine that a transient instruction $x$ affects the execution time of an instruction $y$ that will commit. Even if the commit time of $y$ is unchanged, and no directly visible side channel occurs, if the result forwards to another committed instruction $z$, which then commits earlier as a result, a side channel becomes visible. These transitive effects make it sufficient for \order\ to only maintain that the \textit{commit time} of committed instructions to be unaffected rather than \textit{execution time} per se, as a change in execution time of one instruction can be converted to a change in commit time of another if it has the potential to become visible. Still, to avoid having to consider this transitive impact, it is simpler to design hardware that preserves guarantees over all \textit{execution time}: that is, at no pipeline stage should the effects of a transitive instruction alter the timing of an instruction that may later be committed.

	\myparagraph{Between threads, no such program ordering holds.} For two instructions in two threads $x_{T1}$ and $y_{T2}$,  $x_{T1} \centernot{\xRightarrow{\text{S}}} y_{T2}$ and $y_{T2} \centernot{\xRightarrow{\text{S}}} x_{T1}$ may both hold. Indeed, there is unlikely to be any reason for an ordering outside a guarantee of commit: timing information from $y_{T2}$ can typically only be transmitted to $T1$  once the $y_{T2}$ instruction becomes non-speculative.

	\begin{figure}
		\includegraphics[width=\columnwidth]{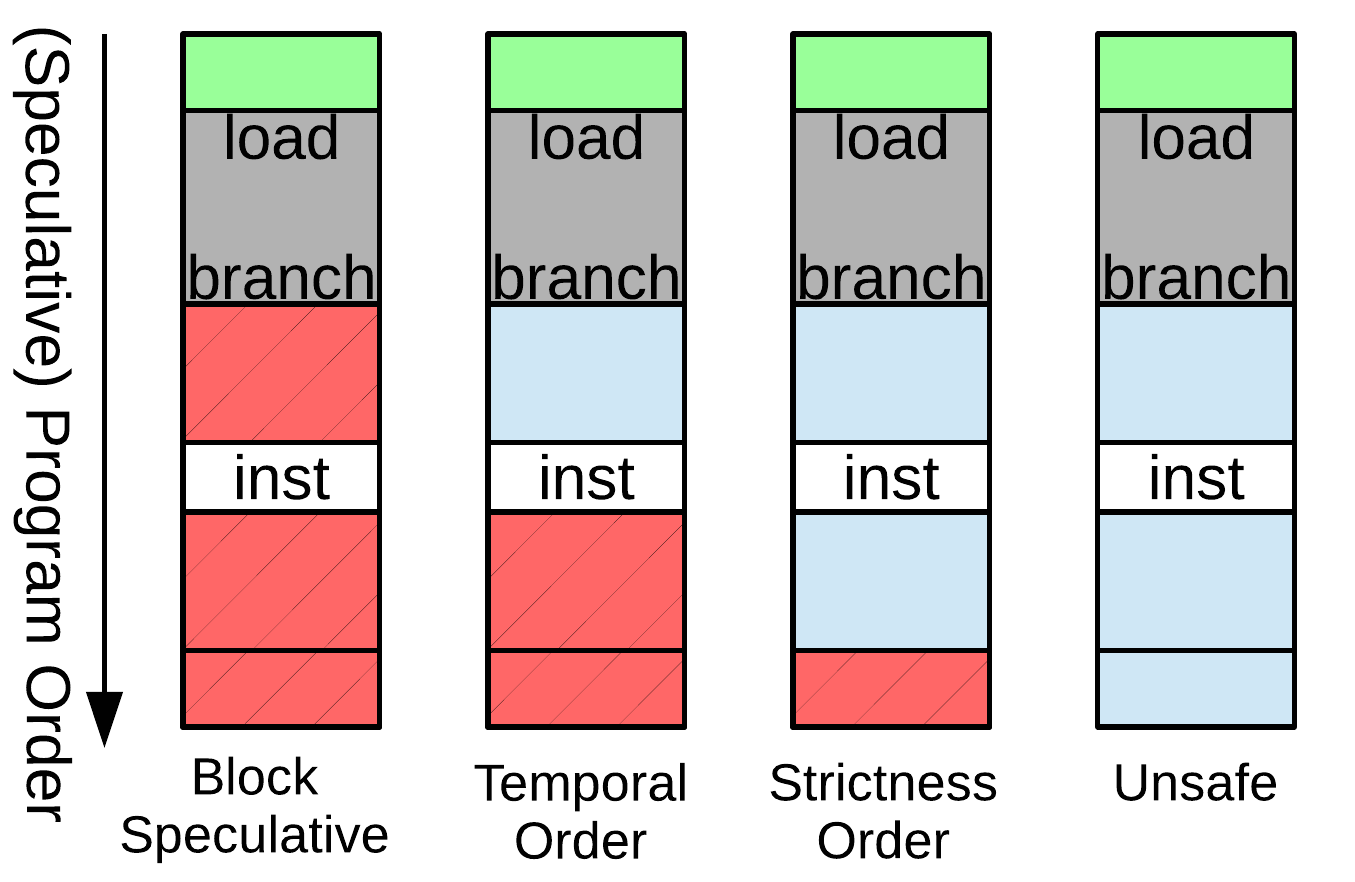}
		\caption{An example of ordering restrictions. Instructions in green are non-speculative, as are those in grey, which are held up by the load's cache miss. Instructions following the grey branch are speculative. Instructions in red (with hatching) cannot affect the timing of the instruction in white before becoming non-speculative, whereas instructions in blue may do so: under Strictness Order, any load before the blue branch can forward data to, or evict data seen by, the instruction in white, or contend for the same resource. \response{By contrast, preventing speculation (Block Speculative) disallows many valid flows, whereas a baseline Unsafe system allows transient execution attacks, by making the effects of operations that do not commit, and can access data disallowed by correct execution, visible to other instructions.} 
		}
		\label{fig:channeldiagram}
	\end{figure}
	
	The result is a preorder relation\footnote{\order\ is named after, and closely related to, strictness analysis~\cite{10.5555/647324.721526}, an optimisation in software to transform call-by-need into call-by-value. A function $f$ is strict in an input $x$ if the function fails to terminate whenever the input fails to terminate. In the same way, an instruction $y$ can strictly observe an instruction $x$ under \order{} if whenever $x$ fails to commit, $y$ fails to commit. 
	}
	, demonstrated in \cref{fig:channeldiagram},  
	which allows significantly freer execution than many previous techniques~\cite{STT-micro,NDA-micro,SDVP,CondSpec}, without affecting security. 
	
	Outside of its basic definition, for the \order\ property to be enforceable, two other properties are assumed.
	
	First,  speculative execution can only be measured by timing impact on at least one valid instruction. Outside of access to advanced performance counters, this is typically true. 
	Second, if  $y \centernot{\xRightarrow{\text{S}}} x$, then as well as $x$'s execution time, $y$ must also not be able to affect \textit{whether} $x$ commits. A speculative technique, that assumed $y \xRightarrow{\text{S}} x$ and rolled back and re-executed $x$ if incorrect, would affect the timing of the program based on the behaviour of $y$, an instruction that was never committed. 
	
	

	\order ing is sufficient to comprehensively prevent any transient execution attack, regardless of whether the effects are timed at any point after correct execution has restarted~\cite{behnia2020speculative,Kocher2018spectre,Lipp2018meltdown} or whether the observed effects are entirely concurrent contention and do not persist past the return to correct execution~\cite{behnia2020speculative}. \response{This is because no misspeculated operation can ever affect the execution time of any committed instruction at any point.}
	
	

	\begin{frm-def}[\lorder ing]
		Let \emph{x} and \emph{y} denote executed instructions. Under \lorder ing, \response{\emph{y} \emph{temporally succeeds} \emph{x}} ($x \xRightarrow{\text{T}} y$), meaning \emph{x} can impact the execution time of \emph{y}, iff 	
		\[commit (x) \vee seq(x,y)\]
		where $commit (i)$ is true iff instruction $i$ will reach the end of the CPU pipeline without being squashed, or already has, $seq(x,y)$ indicates that \emph{x} occurs before \emph{y} in the program's instruction sequence, and $\vee$ is logical disjunction. \label{def:lorder}
	\end{frm-def}
	\response{A simple overapproximation of \order} is \lorder\ (\cref{def:lorder}). This considers each new instruction as more speculative than the last. This is more lenient than in-order execution, or blocking all speculative computation \response{(Block Speculative in \cref{fig:channeldiagram})}, as it only prevents \textit{information flow} (side channels or otherwise) in reverse. Execution can otherwise still occur out-of-order.
	For typical out-of-order systems, which re-execute instructions past misspeculation, a \lorderh\ microarchitecture obeys \order: \response{since there is a total order on instructions that commit within a thread, $seq(x,y) \rightarrow (commit(y) \rightarrow commit(x))$, and by logical tautology, $commit(x) \rightarrow (commit(y) \rightarrow commit(x))$. }
	Instructions can be treated as speculative until they reach commit, and each new instruction can be treated as more speculative than the last.  \response{\name\ implements \lorder\ for simplicity, and because the overapproximation causes little performance loss even though it rejects some valid orderings.} The difference is shown in \cref{fig:channeldiagram}.

	\section{GhostMinion}
	
\begin{figure}
		\centering
		\includegraphics[width=.9\columnwidth]{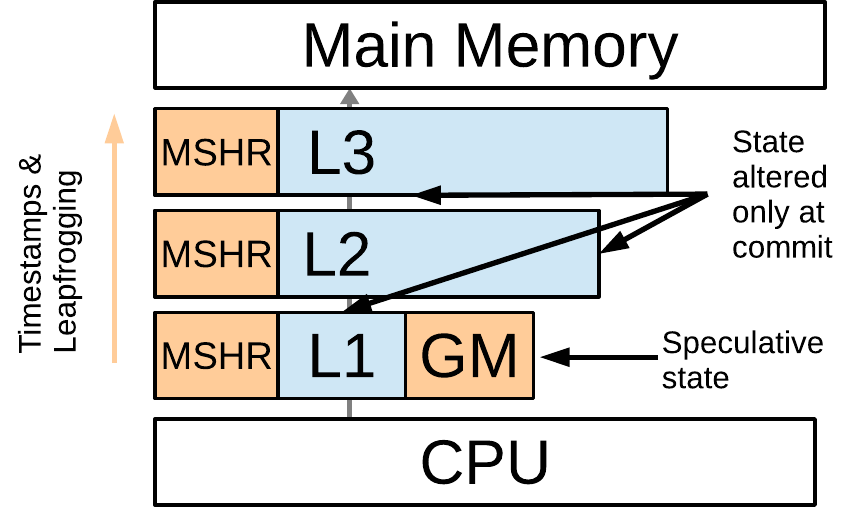}
		\caption{\name\ attaches to each L1 cache, and is designed to cache speculative loads, with \textit{TimeGuarding} to make this data (and any eviction) invisible under concurrent misspeculation. 
			To hide contention, timestamp metadata propagates into the MSHRs at each cache level, allowing loads to be cancelled and replaced (\textit{leapfrogging}).}
		\label{fig:ghostminionsummary}
	\end{figure}

	Here we develop \name, an extension to the cache system and core to enforce \lorder ing (\cref{def:lorder}), and thus \order ing (\cref{def:order}) as a result. A summary is shown in \cref{fig:ghostminionsummary}: we cache the results of speculative memory accesses in a compartment of the L1 cache, labelled a \name. These are only allowed to impact state within the rest of the cache system (the non-speculative cache hierarchy) once an instruction becomes non-speculative. Within the \name, instructions may be forbidden from seeing eviction or insertion of others, depending on their \order\ (\cref{fig:commit,fig:timestamp}).
	The enforcement of this is guaranteed by \textit{TimeGuarding}, which directly translates \lorder{} into hardware.
	Transient contention in the rest of the system, such as the \response{miss-status handling registers} (MSHRs) or non-pipelined instructions, is also subject to TimeGuarding, or when more appropriate \textit{leapfrogging} (\cref{fig:leapfrog}), which allows stealing of resources, followed by replay of the offending operation, to prevent breaking \order\ under greedy out-of-order resource allocation. 
	
	\subsection{How to Achieve \order ing}
	
	Suppose we have two instructions, $x$ and $y$, where $x$ is a misspeculated instruction transmitting via side channel to $y$, which has, or will be, committed. There are three cases: \response{$y$ commits before $x$ enters the pipeline, $y$ is issued after $x$ is flushed, or both are in the pipeline concurrently}.
	
	The first case is trivial. If $y$ commits before $x$ enters the pipeline, then no timing information from $x$ could ever reach $y$. The second is also relatively simple: we must clear any state changes that were caused by misspeculation before restarting correct execution. In order to not affect the timing of future committed instructions, thus violating \order, this rollback to a misspeculation-free state must be timing-invariant to the content, or \textit{volume} of content, that needs to be cleaned up. This means that allowing misspeculation to propagate through a cache system, followed by (serially) undoing~\cite{Saileshwar:2019:CUA:3352460.3358314,wu2020reversispec} the results would violate \order, resulting in a  side channel. Instead, we must make this rollback time-invariant~\cite{Ainsworth_2020} and efficient, by limiting speculative changes to a small region of SRAM (the \name).
	
	The final case, where $x$ and $y$ execute concurrently, is that exploited by techniques such as SpectreRewind~\cite{fustos2020spectrerewind} and Speculative Interference~\cite{behnia2020speculative}. Since both are in the pipeline simultaneously, and $y$ will be committed but $x$ will not, we know that $y$ is before $x$ in (speculative) program order, and that  $y  \xRightarrow{\text{S}} x$ but $x \centernot {\xRightarrow{\text{S}}} y$. Still, $y$ may be delayed from being a candidate for execution, and so $x$ may execute first. To enforce \order, $x$'s execution must not be visible by timing $y$. 
	Not only must we keep speculative resources confined to a \name: 
	we must also prevent any \textit{backwards-in-time} channels before misspeculation is detected. Under \lorder, which treats instructions as a sequence, older instructions in program order must not be able to read data brought in by logically newer instructions, and so must observe a cache miss in the \name\ (\cref{sfig:read}). Newer instructions also cannot evict the data of an older instruction, and so a fill must fail if there are no storage options left within a \name\ cache set that do not violate this constraint (\cref{sfig:fill}). Resource contention must also be hidden, and so older instructions can steal contended resources taken by newer instructions, which must then repeat (\cref{fig:leapfrog}).

	
	
	
	\subsection{Overview}
	
	\name s are placed next to the L1, and accessed in parallel with the L1. 
	They buffer speculative accesses from the rest of the memory hierarchy, which cannot be altered by speculative evictions or fills. The \textit{non-speculative} cache hierarchy may follow any inclusion or exclusion policy, but must be non-inclusive non-exclusive with respect to the \name\ to avoid information leakage.
	
	The non-speculative cache hierarchy never sees state changes from speculative instructions, and so no rollback is required for them on misspeculation. Only the \name\ is wiped on misspeculation, and only it needs restrictions on its data forwarding (\cref{fig:timestamp}). Changes in the non-speculative cache hierarchy can only occur through non-speculative memory accesses (such as the head of the reorder buffer), through a committed instruction moving data out of the \name\ (\cref{fig:commit}), or through the actions of a prefetcher. Still, \name s share the L1's miss status handling registers (MSHRs), and use the L2 and L3's MSHRs, which though they are transient, can cause contention-based timing flow without care. To avoid this, we propagate \lorder\ throughout the MSHR hierarchy, in particular using leapfrogging (\cref{fig:leapfrog}) to steal resources back when the existence of a partially complete memory access would leak. \name s\ cannot cause deadlock or livelock, as all valid instructions become non-speculative, and thus globally visible, once they reach the head of the reorder buffer.
	
	
	

	
	\name s are wiped on misspeculation, in a timing-invariant way, to clear all timestamps above the point of misspeculation\footnote{Unlike in MuonTrap~\cite{Ainsworth_2020}, we do not clear \textit{all} data from the \name, as this could violate \lorder\ if the discovered misspeculation is \textit{itself} speculative (e.g. because a misspeculated branch $C$ relies on a yet-to-be-resolved branch $B$).}. This is done in a single cycle using parallel registers~\cite{Ainsworth_2020}, separate from the L1 cache, which indicate validity and timestamp. \name s do not hold dirty data, and so no writeback is required on this invalidate operation. In addition, to prevent \lorderh\ violation from concurrent execution, constraints on forwarding and evicting data in a \name\ must hold: these are explored below.
	
	
	
	\subsection{Free-Slotting}
	\label{sec:freeslot}
	
	\begin{figure}[t]
		\centering
		\includegraphics[width=.7\columnwidth]{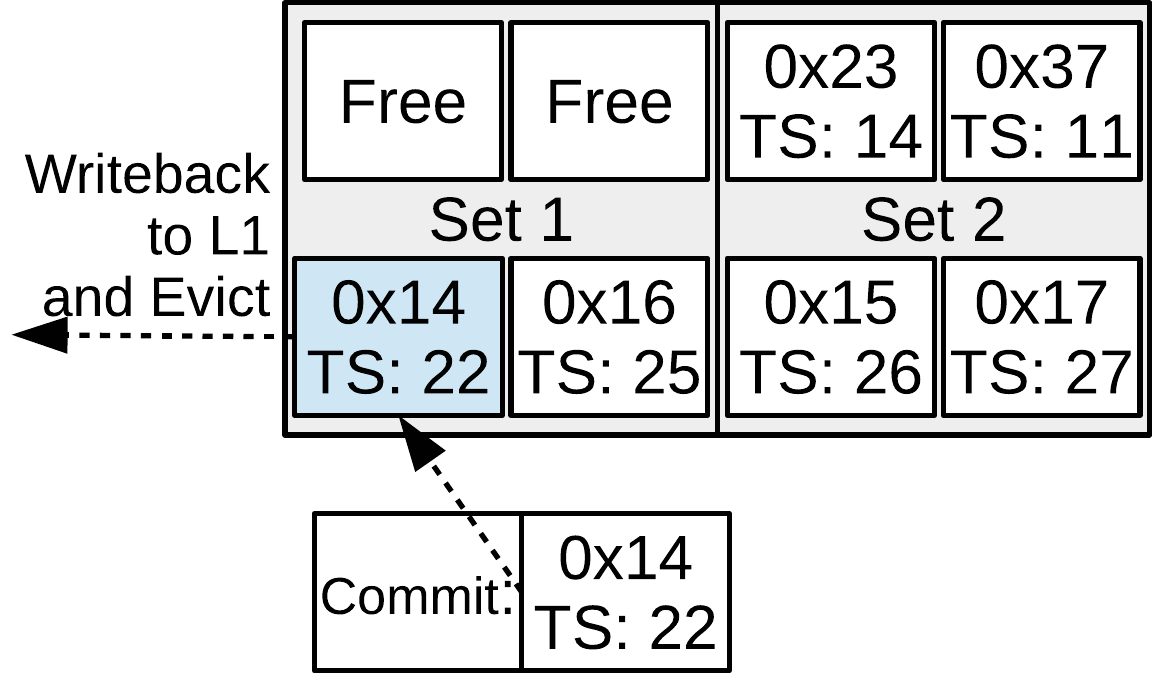}
		\caption{On commit of a load, the \name\ is checked for any cached line that matches, and can be validly read in \order\ by the committed instruction based on its timestamp (TS). If found, it is written to the L1, and evicted from the \name, leaving a ``free slot" for future speculative loads to fill without evicting non-speculative data.}
		\label{fig:commit}
	\end{figure}

	A speculative overwrite (by an instruction $x$), within the \name{}, of data brought in by an instruction $z$ that has committed, is forbidden, as this would result in a backwards-in-time channel by evicting the non-speculative data. Precisely, a speculative instruction $y$, where $z \xRightarrow{\text{T}} y \xRightarrow{\text{T}} x$, but $x \centernot {\xRightarrow{\text{T}}} y$, may then try to read the original, now-evicted cache line from $z$. 
	What follows is that, in order to bring data into a \name, we can only overwrite free slots or data that is \textit{more speculative} than the current load. To avoid resource starvation, when we write a cache line into the non-speculative L1 on commit, we must also evict it from the \name{} to create a \textit{free slot} for speculative memory accesses (\cref{fig:commit}). If the cache system guarantees either exclusion or inclusion for the non-speculative hierarchy, then the line is kept in the \name{} until the data can be moved to the L1 while preserving this invariant, to service requests while exclusion or inclusion are enforced in the hierarchy.
	
	\textit{Free-slotting} is why \name s are placed next to L1 caches, rather than closer to the CPU as in techniques such as MuonTrap~\cite{Ainsworth_2020}. Since a \name\ does not store committed data, it no longer makes sense to use a \name\ as a fast L0 cache for frequent lookups: instead, it should be accessed in parallel with the  L1. 
	This writeback-on-commit is the only way data can escape to the outside world: otherwise, it is thrown away, to avoid affecting other caches\footnote{Nevertheless, the state of the non-speculative cache hierarchy can validly be different from a sequential execution, but only in ways that do not violate Temporal Order. For example, if instruction $y$ at timestamp 43 fills the GhostMinion, then $x$ at timestamp 42 fills to the same slot, which it sees as free to preserve Timestamp Order. This means that $y$, never fills to the L1 despite both being valid. This is permitted because, while $x$ cannot see the out-of-order effects of $y$, $y$ is validly permitted see the out-of-order effects of $x$. Thus, maintaining equivalent state to an in-order execution is not necessary to maintain \lorder\ or \order.}.

	\subsection{TimeGuarding}
	\label{sec:timestamp}
	
	\begin{figure}[t]
		\begin{subfigure}{\columnwidth}
			\centering
			\includegraphics[width=.9\columnwidth]{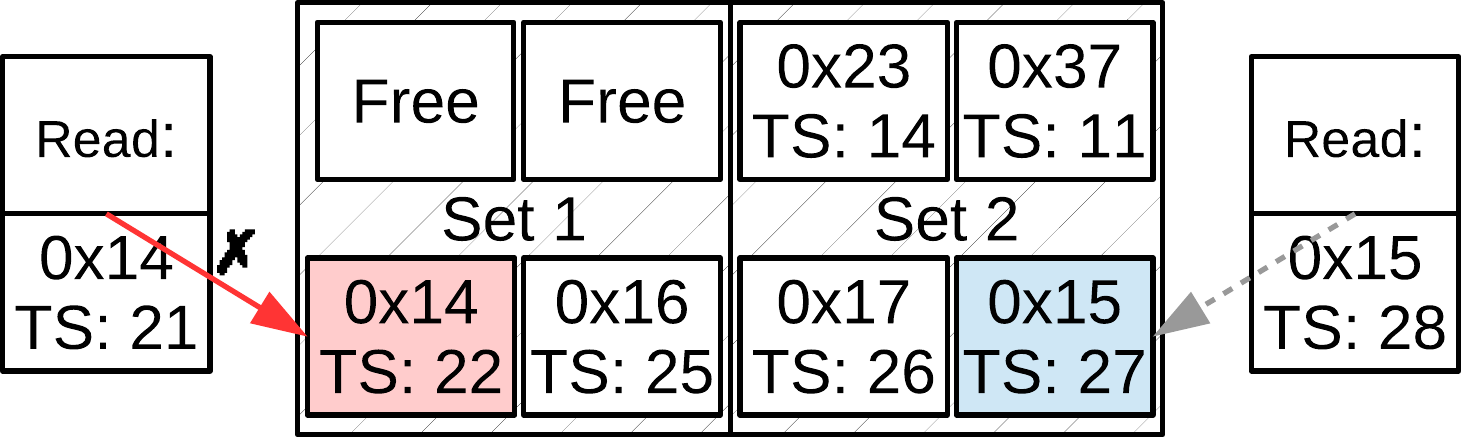}
			\caption{Reads only succeed if the timestamp of a cache line is less than or equal to the timestamp of the instruction. In this example, if the load of \textit{0x14} at timestamp $22$ did not commit, but the same load at timestamp $21$ committed and observed $22$'s cache line, this would break \order. For the load of \textit{0x15}, the timestamp $28$ is higher than $27$; the former cannot commit if the latter fails, so the channel is permitted.}
			\label{sfig:read}
		\end{subfigure}
		\begin{subfigure}{\columnwidth}
			\centering
			\includegraphics[width=.7\columnwidth]{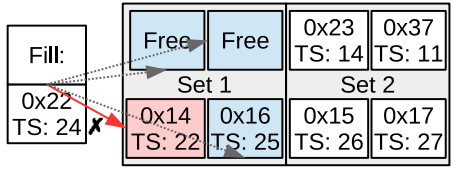}
			\caption{Filling the \name{} is the opposite: it can only overwrite data with a greater-than or equal timestamp (or a free slot). 
			}
			\label{sfig:fill}
		\end{subfigure}
		
		\caption{To enforce \order, \name\ puts TimeGuarding restrictions on which data can be read or overwritten, to make the existence of concurrent misspeculation invisible. This prevents any backwards information flow, from existence or non-existence of data, and works regardless of the associativity of the \name. }
		\label{fig:timestamp}
	\end{figure}
	
	To track \lorder, each issued instruction is assigned a timestamp. This is attached to speculative memory accesses of the \name; a speculative load is only allowed to \textit{read} a value from the \name\ if its timestamp is \textit{greater or equal} than the value in the \name\ (\cref{sfig:read}).
	
	A cache \textit{fill} in the \name{} is only allowed if \textit{a)} the block is free, or \textit{b)} the new request has a \textit{lower} timestamp than the block it wishes to evict (\cref{sfig:fill})\footnote{Any policy that does not reveal the difference between \textit{a)} a ``free slot'' and \textit{b)} a location taken by a block at a higher timestamp, to an instruction at a lower timestamp, is valid; we choose to take free slots if available, and evict the highest-timestamped available location in a set otherwise. This is because only the highest-timestamped instruction is aware that the \name{} is in fact full.}.  Whether a suitable space is available or not is dictated by the value of a given address, assuming the structure is not fully associative, and the number of sets available. If no sets can validly be used within a given line without violating TimeGuarding, the load is returned to the CPU without being written to the \name: it will thus not be available to write back to the L1 on commit.
	
	
	
	Logically and behaviourally, the TimeGuard can be considered a timestamp increasing to infinity. Implementation-wise, the maximum timestamp is sized to be twice the number of reorder-buffer entries, as a sliding window\footnote{Since there will only ever be $N$ maximum instructions in the reorder buffer at once, allocated in timestamp order, each individual instruction can only execute concurrently with $2N$ instructions. It is sufficient to allocate timestamps in-order with a wrap-around at $2N$, and gives the constraint that an instruction at timestamp $t$ can only overwrite a cache line at timestamp $t .. ((t+N) \mod (2N))$, and only read all others. 
	}. 
	
	\subsection{Leapfrogging}
	\label{ssec:leapfrog}
	
	\begin{figure}[t]
		\includegraphics[width=\columnwidth]{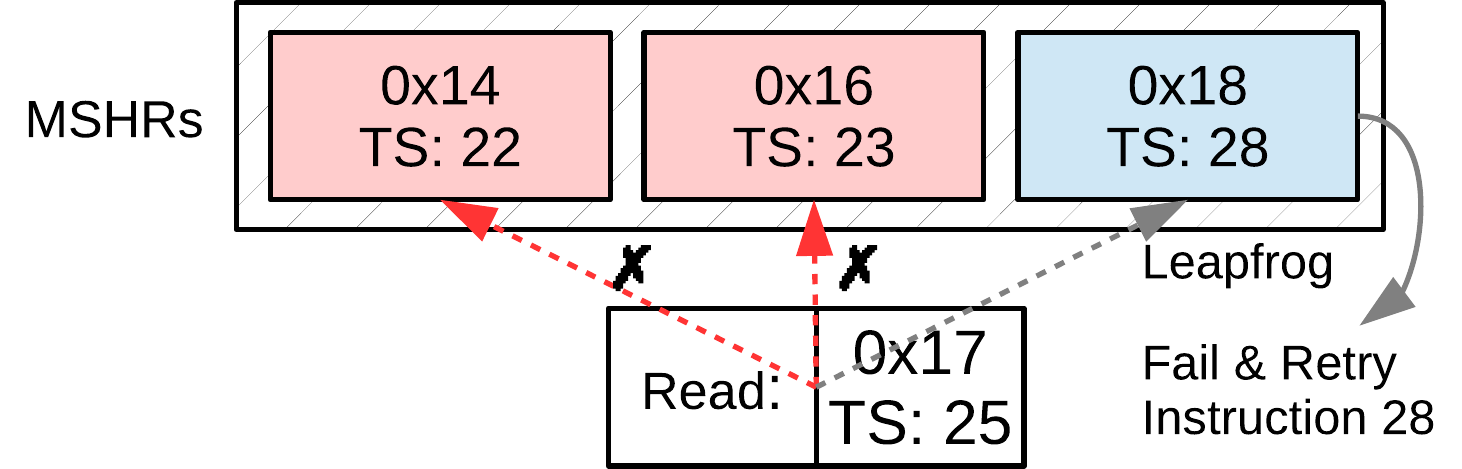}
		\caption{An example of \textit{leapfrogging}. Here, no miss status handling registers (MSHRs) in the cache are free. The read request we are attempting has a timestamp of $25$, and so cannot affect the execution of the MSHR requests at timestamps $22$ and $23$. The timestamp $28$ must be invisible to our current request: the new read, at timestamp $25$, replaces the one at timestamp $28$, which fails and must reattempt later.
		}
		\label{fig:leapfrog}
	\end{figure}

	\name{} allocates and schedules resources to instructions greedily. A speculative instruction $y$ may be allocated the last free MSHR in the cache, when an instruction $x$, before $y$ in program order, becomes available to execute. If $y$ then delayed the execution of $x$, this would be a violation of \lorder{} (and thus \order{}), and so to fix this, $x$ must be able to steal $y$'s MSHR from underneath it, in a process called \textit{leapfrogging} (\cref{fig:leapfrog}). This causes the load for $y$ to lose its MSHR resource; the core treats it as though the MSHRs are full, and so the load must be reattempted once new resources are available, to hide its execution from $x$.
	To implement this, a thread's load requests' timestamps are propagated up the entire MSHR hierarchy: a load can fail due to leapfrogs in any MSHR up to main memory.
	
	

	A special case of leapfrogging, which we call \textit{timeleaping}, is when instructions $x$ and $y$, such that $x \xRightarrow{\text{T}} y$  but  $y \centernot {\xRightarrow{\text{T}}} x$, execute concurrently, with $y$ issuing first, and both attempt to bring in the \textit{same} cache line. Here, they can both be attached to the same MSHR, and so $y$ need not be cancelled and replayed\footnote{If $x$ literally takes the same MSHR, and thus prioritises the taking of a \textit{particular} MSHR based on whether $x$ and $y$ match, then leapfrogging should otherwise occur on the highest timestamped MSHR to avoid breaking \lorder. If we instead used a random eviction policy of MSHRs for any timestamp above $x$'s, an instruction $z$ using another MSHR, which believed there was still free space, could learn that the addresses of $x$ and $y$ match even if  $x \xRightarrow{\text{T}} z \xRightarrow{\text{T}} y$ and   $y \centernot {\xRightarrow{\text{T}}} z \centernot {\xRightarrow{\text{T}}} x$, if it were therefore never leapfrogged.}. 
	To hide the timing effect, the load is still leapfrogged: it is restarted at each cache. As with standard leapfrogging, this may cause \textit{cascading leapfrogs}, in this case of the same request, in multiple different cache levels. This achieves the same timing response as if only $x$ executed the load, which since  $x \xRightarrow{\text{T}} y$ may validly forward the result to $y$.
	
	
	
	\subsection{Coherence}\label{ssec:coher}
	
	Timing data can also leak through influencing coherence protocol states~\cite{SpectrePrime}. Between threads, no \order\ exists in general (\cref{sec:order}). This means that coherence states, either in non-speculative caches or in other \name s, must not be altered until an instruction commits. 
	
	\name{} uses a technique taken from MuonTrap~\cite{Ainsworth_2020}, which is sufficient to avoid altering non-speculative coherence state, and avoid observing timing effects from speculative state. A block within a \name\ can only be in Shared or Invalid\footnote{The ReadExclusive memory request necessary for stores can only occur once the store becomes non-speculative~\cite{Ainsworth_2020}. Stores already typically occur at commit, since they are challenging to roll back, and do not sit on critical paths unlike loads, due to the store buffering permissible in conventional \response{TSO and relaxed consistency models}.
	}. A \textit{Shared} block can only exist in a \name{} provided no \textit{Exclusive/Modified} blocks sit in non-local parts of the cache hierarchy; that is, caches not along a direct path from the CPU to Main Memory, for example in a private (non-speculative) cache of another core. Loads that would violate this wait until non-speculative, filling directly into the L1, before they can gain a coherent copy.
	
	Equivalently, if a line exists in exclusive or modified in a localised part of the non-speculative cache (for example, a private L2 or L1), a valid copy cannot exist, in a \name\ or in a non-speculative cache, down any other part of the hierarchy. This means any upgrade (for example, to perform a store) need only invalidate \name s on paths \textit{closer to a CPU} than such a copy, in a method timing-insensitive to the speculative contents of a \name~\cite{Ainsworth_2020}.
	
	Outside the coherence protocol itself, \name\ can forward a speculative non-coherent copy of a value, in Exclusive or Modified in a non-local part of the cache hierarchy, to the \name{}. On an attempt to commit an instruction using this non-coherent copy, the load is replayed non-speculatively, and the instruction itself re-executed if the value has subsequently changed. This technique is taken from InvisiSpec~\cite{yan2018invisispec}, which uses it for all speculative loads. If the load originally received a valid shared copy, rather than a value-predicted version, and can thus commit without this check, and no shared copy existed in any non-local part of the non-speculative hierarchy, the L1 cache may then issue a decoupled upgrade~\cite{Ainsworth_2020} to move the data into exclusive.
	
	\subsection{Prefetching}
	
	To avoid information leakage, prefetchers in non-speculative parts of the cache hierarchy can only be trained on non-speculative input. 
	To achieve this, speculative memory accesses are tagged with the cache level they were brought in from. On commit, not only is the value (if relevant) brought in to the \name, committed to the L1, a prefetcher notification is also sent to the level it was brought in from (and any levels closer to the CPU), provided it has a prefetcher at that level. Again, this technique is inspired by MuonTrap~\cite{Ainsworth_2020}.
	
	Prefetchers can only be trained using data from speculative streams if they prefetch into the \name. For prefetchers close to the CPU, such as fetch-directed instruction prefetching~\cite{FDIP}, this may be desirable. These prefetches are timestamped to the value of the last instruction used as input to the prefetcher: only instructions with equal or higher timestamp can observe the prefetches. 

	\subsection{Instruction \name}
	
	The instruction cache can also be affected by transient execution attacks. These can be caused in two ways. Branches can be speculatively resolved not only by lookup in the predictor and branch target buffer, but also as a result of computation (for example, a branch or branch target conditional on the value of a secret). Alternatively, secrets can generate a contention effect on instruction-fetch progress.
	
	
	Because instruction fetch is executed in (speculative) program order, protection is easier than for the data hierarchy, since no instruction reordering can occur, so for a pipeline's concurrently executing instructions, \lorder\ cannot be broken via passing information \textit{backwards-in-time}. We need only worry about leaking information from the end of misspeculation to the start of re-execution.
	Still, we must be able to return to a cache state unaffected by wrong speculation, and in a timing-invariant way that does not reveal how much state was changed. 
	This requires being able to wipe an instruction \name\ with a parallel invalidate operation, like for a data \name. 
	Further, if $y \centernot{\xRightarrow{\text{T}}} x$, then $y$ cannot evict $x$'s cache line unless $x$ has been committed. We handle this via TimeGuarding (\cref{sec:timestamp}), and in practice, the instruction \name\ is similar to the data \name\ mentioned above, without coherence requirements.
	
	
	\subsection{Remaining Channels}
	
	Though for a system to truly be transient side-channel free, the entire system must obey \order, the techniques here are sufficient to prevent load forwarding, either forwards or backwards in time, being used to extract such an attack. Further, since the cache system is the most complex holder of speculatively updated soft state, it is responsible for the vast majority of overheads in mitigation. We argue here that for other system components, the overhead of providing \order\ is likely to be negligible, and that we can reuse techniques developed here or elsewhere.
	
	\myparagraph{Within-core structural hazards} \order\ can be violated by a resource anywhere within the core being taken by a misspeculated instruction, delaying the scheduling of an instruction that will be committed. For single-cycle or pipelined operations, the solution is simple: we can prioritise instructions with lower timestamp. 
	For multi-cycle operations that can cause structural hazards~\cite{fustos2020spectrerewind,DataOblivious,behnia2020speculative} 
	scheduling can be done in \order\ via TimeGuarding. Alternatively, scheduling could be done optimistically, with leapfrogging used to cancel operations if \order\ is violated. We implemented the former in our gem5 simulation, such that functional units that are not pipelined (in our case, the IntDiv, FloatDiv, and FloatSqrt units) may only be issued a speculative operation once all previous speculative operations in timestamp order, that may use the same unit, have issued. This means that all IntDivs, for example, are executed in (speculative) program order with respect to each other. This causes no non-negligible slowdown in any workload we evaluated (maximum 0.08\%). In fact, we saw a slight speedup on some workloads (4\% geomean on SPEC CPU2006, and 2\% on SPECspeed 2017). This is unsurprising; non-pipelined functional units are rarely part of address calculation for loads, and so are unlikely to be improved by long-scale reordering, and by favoring the scheduling of earlier operations, we cause entries to leave the reorder buffer more quickly. Since this shows a speedup, not a slowdown, we do not include this cost saving in the rest of the evaluation, to avoid masking overheads of other mitigations.
	
	\myparagraph{Load-Store Queue} Here, protection is simpler than for the cache; it is naturally designed to transmit data in forwards-program order, thus providing \lorder\ for data flow. Still, to protect against the opposite (data not flowing when it otherwise should, via contention), TimeGuarding (\cref{sec:timestamp}) and leapfrogging (\cref{ssec:leapfrog}) can be used.
	
	
	\myparagraph{Transient contention from multiple threads} Because no total ordering exists for concurrently executing instructions on multiple threads, leapfrogging cannot be used to hide contention between them, particularly on MSHR resources in shared caches. Data-oblivious execution~\cite{DataOblivious} presents one solution to this, where each instruction has a predicted use of such resources, and must repeat once non-speculative if this prediction is incorrect. Still, the concept of \order\ makes this problem much simpler. As it provides a method of avoiding timing leakage in the wrong order within a thread, it vastly simplifies the prediction complexity. Instead of needing to predict \textit{for every instruction}, \order\ makes predicting resource utilisation \textit{per thread} sufficient to avoid leakage. This allows simple macro-level allocation based on recent committed behaviour to be used for shared resources. For simultaneous multi-threaded systems, cross-thread instruction port contention~\cite{Smother} can be prevented similarly.
	
	
	\myparagraph{Address translation} \name s should also be attached to TLBs~\cite{Ainsworth_2020} and page table walker caches. Behaviour is similar to those developed here, without coherence protection.
	
	\myparagraph{DRAM contention} Speculation-hiding techniques in the literature disallow speculative DRAM accesses entirely~\cite{DataOblivious} or keep them out of scope~\cite{Ainsworth_2020,yan2018invisispec}. 
	The biggest issue is open-page policies acting as an implicit cache. Still, allowing only non-speculative accesses to leave pages open may be feasible, since they are most useful with the spatially local accesses from prefetchers rather than the out-of-order core.
	
	\myparagraph{Other soft state:} There are many structures that store soft state within systems, such as branch predictors, branch target buffers, and even frequency-voltage scaling~\cite{NetSpectre}. Unlike with the cache, we believe there is little need to update these with speculative values, and so \order\ should in general be achieved by only updating them non-speculatively.

	\subsection{Optimizations}
	
	This proof-of-concept microarchitecture, implementing \lorder, leaves \response{potential optimisations} on the table.

	\myparagraph{Early Commit} Instead of treating instructions as speculative until commit (as in \name, MuonTrap~\cite{Ainsworth_2020}, InvisiSpec-Future~\cite{yan2018invisispec} and STT-Future~\cite{STT-micro}), instructions could be treated as non-speculative provided their branches are resolved (as in InvisiSpec-Spectre~\cite{yan2018invisispec} and STT-Spectre~\cite{STT-micro}). This requires some more tracking within the processor, and stops any inherent protection against exception attacks~\cite{Lipp2018meltdown}, but these may be dealt with via other means~\cite{kaiser}.
	
	\myparagraph{Full \order} The \lorder\ that \name\ implements is, though simple, more restrictive than necessary. To support \order, rather than labelling each individual instruction with a new timestamp, this could instead be done by assigning a new timestamp after each speculatively predicted branch instruction (or more generally, operation that has the potential to fault). To extract further performance, groups with differing timestamps could be merged when intermediate speculation is resolved.
	
	

	
	\subsection{Summary}
	
	Here we have presented \name, a cache system designed to obey \lorder, a simple version of \order. To avoid any information leakage to concurrent instructions that violate this order, we have introduced a variety of techniques to make such instructions' behaviour visible when in a permitted order, and invisible otherwise: this is greatly simplified by the fact that \order\ is total for a pipeline's execution, and thus for any two instructions, at least one can see the effects of the other. 
	

	\begin{table}[t]
		\small
		\begin{tabularx}{\columnwidth}{lX}
			\multicolumn{2}{c}{\textit{Core}}\\
			\midrule
			Core & 4-Core, 8-Wide, Out-of-order, 2.0GHz \\
			Pipeline & 192-Entry ROB, 64-entry IQ, 32-entry LQ, 32-entry SQ, 256 Int / 256 FP registers, 6 Int ALUs, 4 FP ALUs, 2 Mult/Div ALU \\
			Tournament & 2-bit, 2048-entry local,
			8192 global,\\
			Predictor\ & 8192 choice, 4096 BTB, 16 RAS \\
			\\
			\multicolumn{2}{c}{\textit{Caches with \name s}}\\
			\midrule
			L1 ICache & 32KiB, 2-way, 2-cycle latency, 4 MSHRs \\
			L1 DCache & 64KiB, 2-way, 2-cycle latency, 4 MSHRs \\
			D/I GhostMinions & 2KiB, 2-way, accessed with I/D cache\\
			\\
			\multicolumn{2}{c}{\textit{Rest of system}}\\
			\midrule
			L2 Cache & 2MiB, shared, 8-way, 20-cycle latency, 20 MSHRs, stride prefetcher (64-entry RPT) \\
			Memory & DDR3-1600 11-11-11 \\
			OS & Ubuntu 14.04 LTS \\
		\end{tabularx}
		\caption{System experimental setup.}
		\label{tab:setuptable}
	\end{table}
	
	\section{Experimental Setup}
	We use gem5~\cite{Binkert:2011:GS:2024716.2024718} to model the high-performance Aarch64 system in \cref{tab:setuptable}. This is based on the default out-of-order setup in gem5 also used in previous Spectre defenses~\cite{yan2018invisispec,STT-micro,Ainsworth_2020}.
	We simulate SPEC CPU2006~\cite{SPEC2006} in syscall emulation mode, fast forwarding for 1 billion instructions, then running for 1 billion. SPECspeed 2017 is fast-forwarded for 10 billion, to account for longer initialization, for workloads that would run in gem5. Parsec~\cite{Parsec} (sim-medium) is run in full-system mode on four threads, to completion, for workloads that compile on Aarch64. Performance breakdowns are performed on SPEC CPU2006, as this shows the highest overhead. Results for InvisiSpec~\cite{yan2018invisispec}, STT~\cite{STT-micro} and MuonTrap~\cite{Ainsworth_2020}, on the same Aarch64 setup, are compared against.

	\section{Evaluation}
	
	The performance overheads of \name\ are minimal for most workloads (2.5\% geomean on SPEC CPU2006, 0\% Parsec, 0.6\% SPECspeed 2017). Though the worst case is 30\%, this is a fundamental to hiding incorrect speculation rather than specific to \name. The prevention of backwards-in-time channels, via TimeGuarding and leapfrogging, rarely affects performance, as such flow is rare.
	
	\subsection{Comparison}
	\label{sec:comparison}
	
	\begin{figure*}[t]
		\centering
		\includegraphics[width=2\columnwidth]{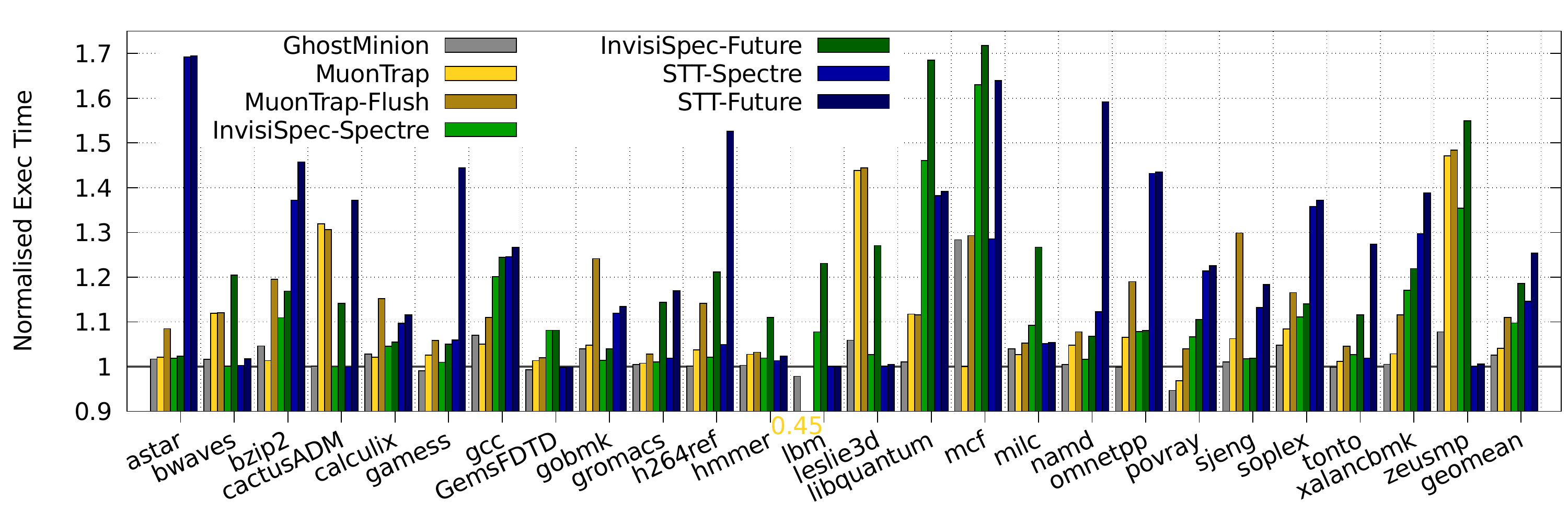}
		\caption{The performance overhead of \name\ on SPEC CPU2006, versus various techniques from the literature.}
		\label{fig:versusspec}
	\end{figure*}
	
	\begin{figure}[t]
		\centering
		\includegraphics[width=\columnwidth]{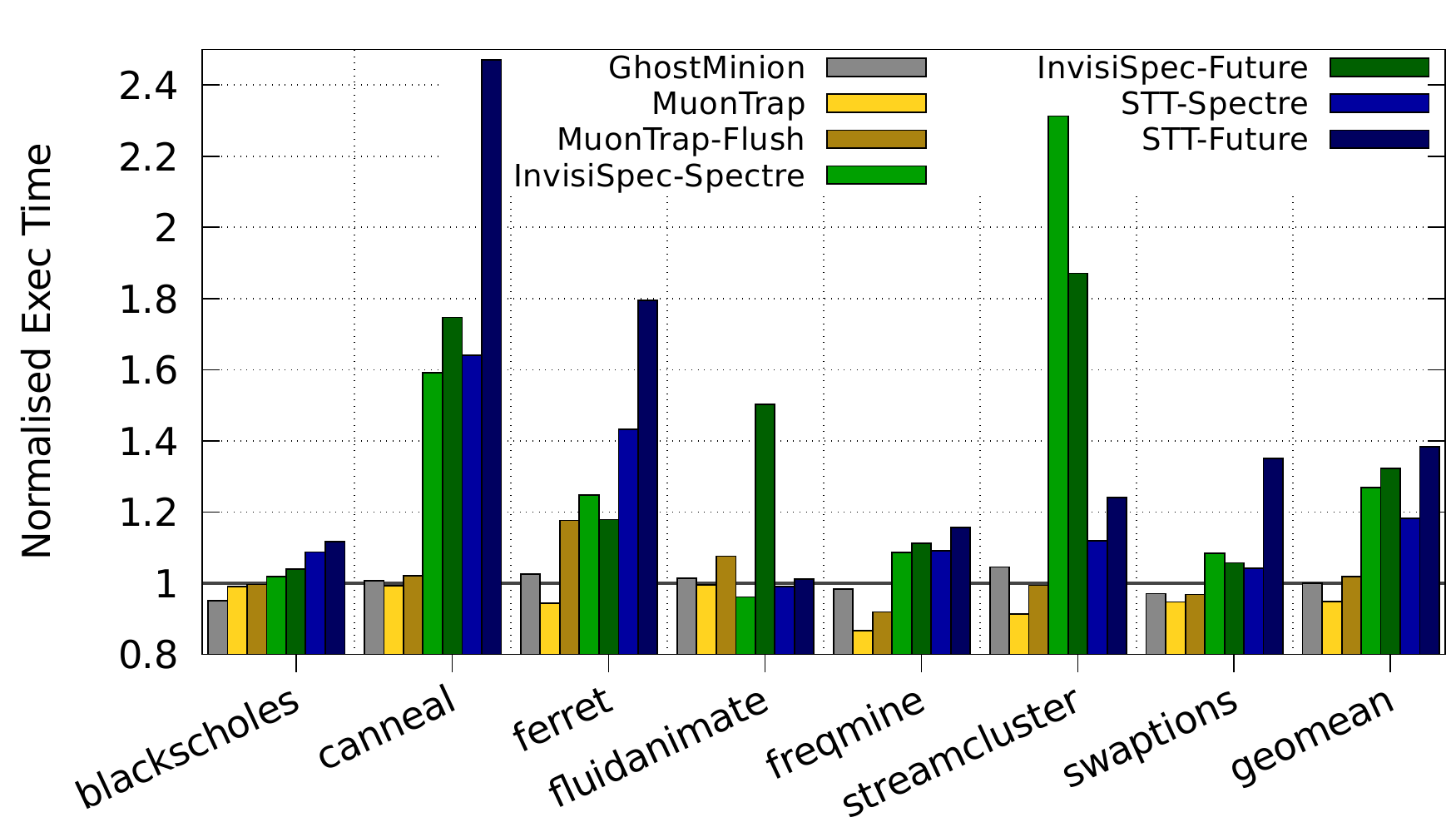}
		\caption{The performance overhead of \name\ on 4-thread Parsec, versus various techniques from the literature.}
		\label{fig:versusparsec}
	\end{figure}
	
	\begin{figure*}[t]
		\centering
		\includegraphics[width=2\columnwidth]{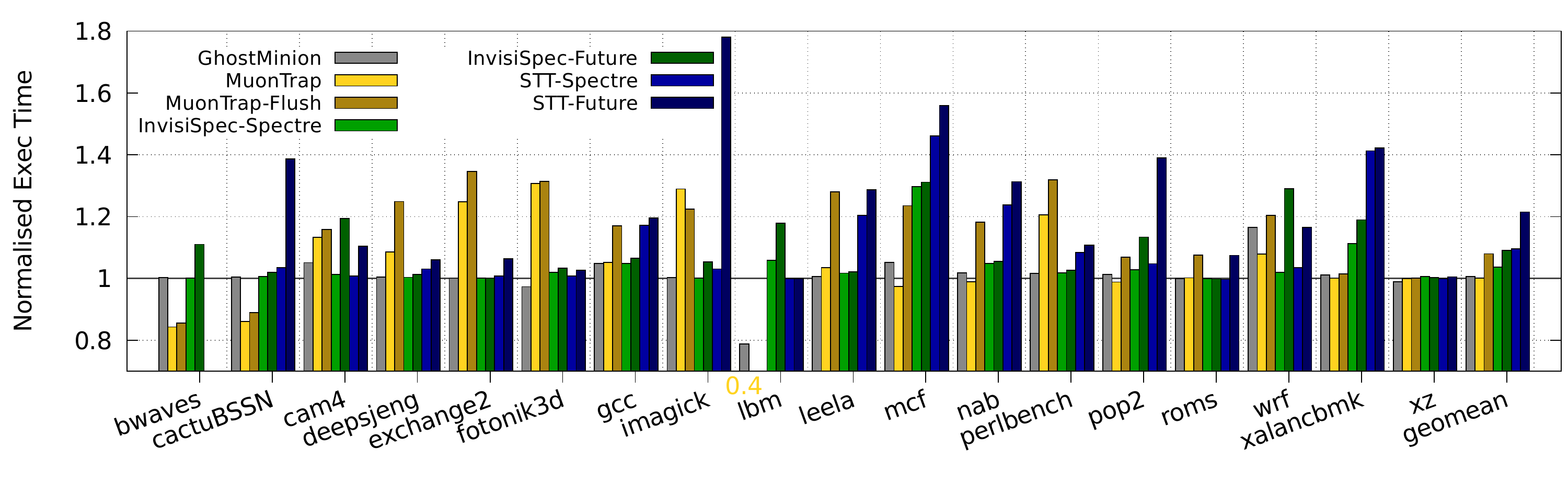}
		\caption{Performance on SPECspeed 2017.}
		\label{fig:spec17}
	\end{figure*}

	\Cref{fig:versusspec} shows the performance of \name\ on SPEC CPU2006 relative to other techniques in the literature. \Cref{fig:versusparsec} shows the same on 4-thread Parsec, and \Cref{fig:spec17} on SPECspeed 2017. While InvisiSpec~\cite{yan2018invisispec} follows a similar strategy of hiding speculative loads (in a Load-Store Queue rather than a cache, and without precise \order\ guarantees, making it vulnerable to Speculative Interference~\cite{behnia2020speculative}), its overheads are much higher: this is mainly down to the coherence protocol requiring a repeated load when an instruction becomes non-speculative before it is allowed to commit, unlike GhostMinion's MuonTrap-like technique~\cite{Ainsworth_2020} which takes such operations off of the critical path. This is despite GhostMinion allowing the use of set-associative structures in a leak-free way via TimeGuarding, and without redundant copies of a cache line for every load-queue entry. GhostMinion follows the same threat model as InvisiSpec-Future rather than InvisiSpec-Spectre, in that all operations are considered unsafe until commit time, rather than until all branches have been decided (but not exceptions). This has little impact on the performance of \name, because few critical-path operations must wait until commit.
	
	The speculation-hiding mechanism of \name, made safe using the \order-based TimeGuarding and leapfrogging, allows more permissive execution of programs than the speculation-prevention mechanisms within STT~\cite{STT-micro}, which only allow potentially-leaking memory accesses, such as those based on another speculative (tainted) load, to occur once all previous branches have been decided. There are many workloads, such as astar, gamess, gromacs, libquantum, namd, omnetpp and xalancbmk, where STT shows large overheads when GhostMinion shows none at all. 
	
	Despite offering stronger guarantees than MuonTrap~\cite{Ainsworth_2020}, a technique that hides speculative memory accesses in an L0 filter cache, rather than a parallel structure next to the L1, \name\ outperforms both MuonTrap techniques on SPEC CPU2006, and outperforms MuonTrap-Flush on Parsec and SPECspeed 2017. MuonTrap is designed primarily as a cross-process defence, though with an optional flush post-misspeculation (MuonTrap-Flush). Even though one of GhostMinion's mechanisms to support \order\ is this same flush, the placement of \name\ next to the L1 cache, with access in parallel, is better suited to frequent flushing than an L0 filter cache is, where the latter is often empty, and thus serves only to increase the access latency of the L1 data/instruction caches. Since a \name\ does not store values already in the attached L1, it can be smaller and less associative than a MuonTrap.

	Some workloads still show significant overhead with a \name. Mcf in SPEC CPU2006 shows almost 30\% overhead. MuonTrap's base system shows no overhead here, whereas STT-Spectre and MuonTrap-Flush show similar overhead to \name, with other techniques higher. The base MuonTrap is the only technique that allows access to data brought in via misspeculation past the point of the processor restarting execution: MuonTrap-Flush and \name\ clear it, InvisiSpec makes it invisible through its Load-Store queue, and STT doesn't execute dangerous loads at all. What this suggests is that mcf, and other workloads such as bzip2, gcc, gobmk, soplex, wrf and zeusmp, rely on this incorrect speculation to do useful prefetching work, and so any safe mechanism must prevent it and lose the relevant performance gains. To improve this, we would require more accurate speculation, or a (speculation-safe) prefetcher able to target complex memory accesses. 
	

	\subsection{Analysis of Overheads}
	
\begin{figure*}[t]
		\centering
		\includegraphics[width=2\columnwidth]{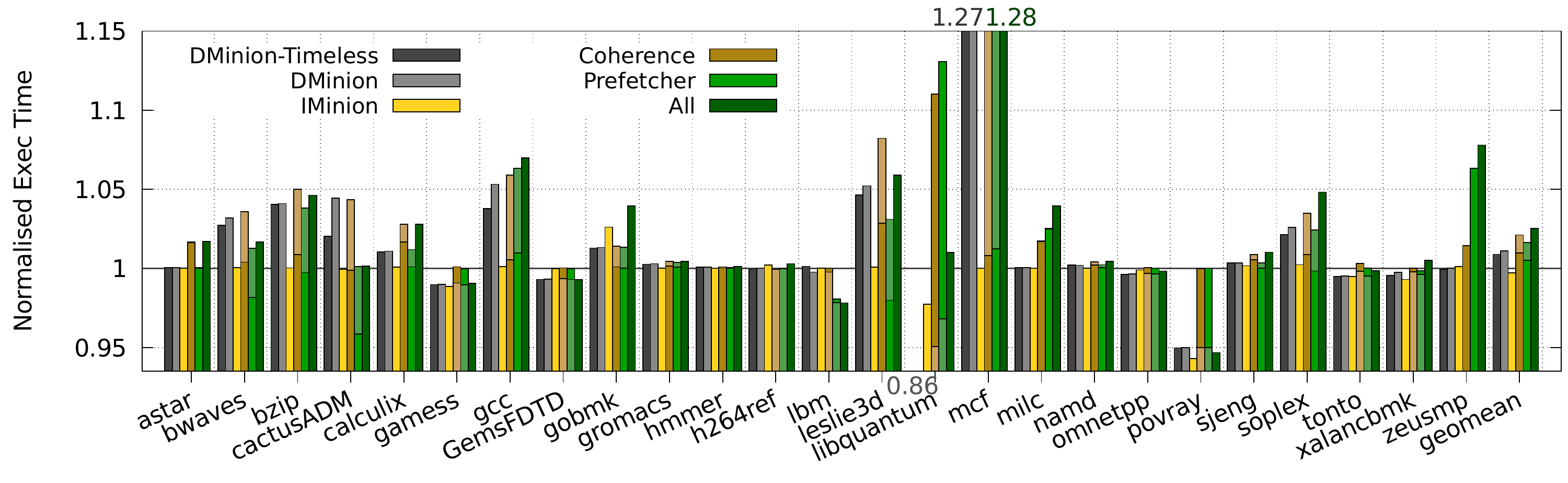}
		\caption{Breakdown of overheads from \name. \response{Dminion-Timeless shows a \name{} with no concept of timestamping, but still wiped on misspeculation, so able to protect against standard Spectre attacks but not backwards-in-time attacks.} The DMinion, IMinion and combined techniques are relative to an unprotected system; for Coherence / Prefetcher, the lighter bars show overhead (with DMinion enabled) relative to the baseline, and the darker bars relative to the DMinion alone.}
		\label{fig:features}
	\end{figure*}

	\Cref{fig:features} breaks down the overheads of each part of the full \name\ on SPEC CPU2006. Here, we see that most of the performance overhead comes from both the data-side \name, and the coherence protocol extensions, with a smaller contribution from the prefetcher (helping lbm by presenting a stabler pattern, and hindering others by affecting timeliness) and none from the instruction side.
	
	Gcc, leslie3d and mcf are hurt by the lack of access to misspeculated data on the data-side, as is wrf in SPECspeed 2017. Still, a DMinion improves performance slightly on workloads such as gamess, GemsFDTD, tonto, povray, and particularly libquantum. This is down to reduced conflict misses, particularly when part of the active set is used repeatedly in the DMinion before it is committed (and thus evicts the conflicting sets).
	Still, this rare speedup is brittle; we see for Libquantum that, if either the coherence or prefetcher mechanisms are added, the improvement disappears. 

	\subsection{TimeGuarding and Leapfrogging}
	
\begin{figure}[t]
		\includegraphics[width=\columnwidth]{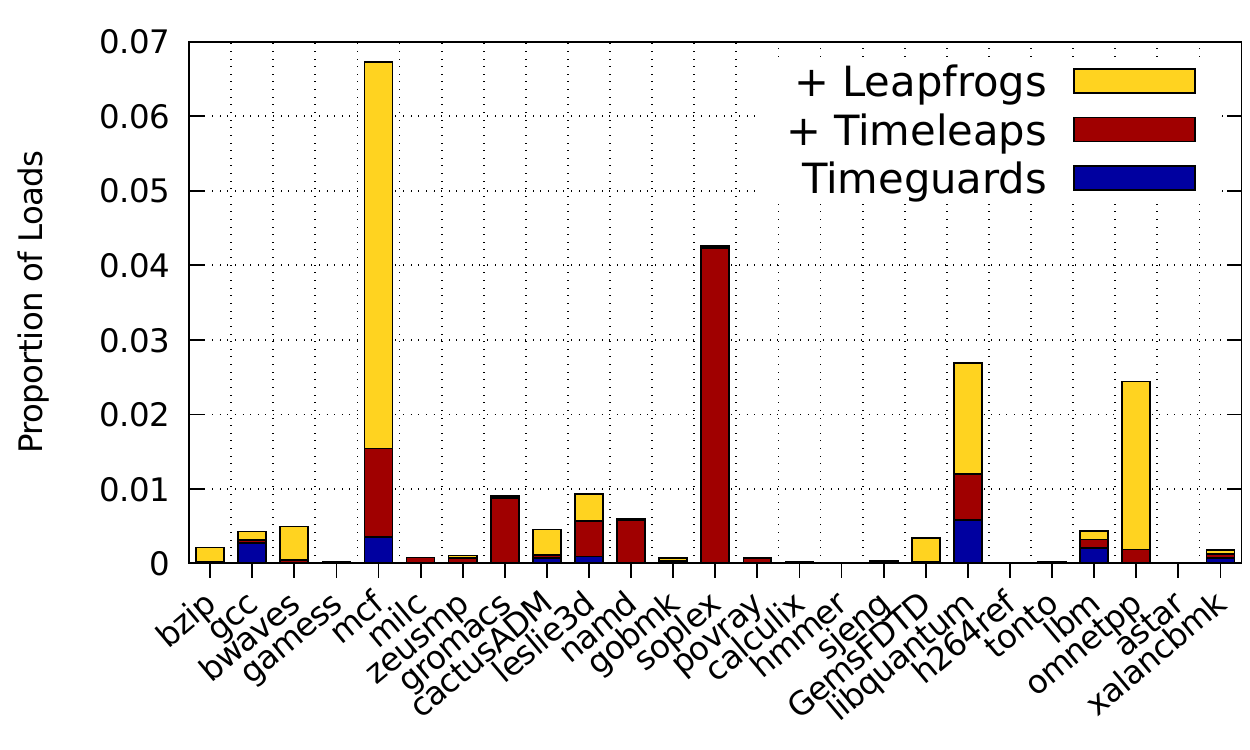}
		\caption{\textit{Backwards-in-time} prevention is rarely triggered.}
		\label{fig:timeguards}
	\end{figure}

	The overheads from protecting against back\-wards-in-time channels within current speculative state, as opposed to lack of resource in the \name\ or clearing on misspeculation, rarely affect benign code. \response{As seen in \cref{fig:features}, the extra overhead is 0.2\%, with a worst case of 2.5\%.} \Cref{fig:timeguards} shows the proportion of memory requests that experience \textit{TimeGuarding} in the \name, to prevent access to a cache line brought in by a future memory request, those that experience \textit{timeleaping}, where a memory request currently in the MSHRs would deliver data backwards in time, and \textit{leapfrogging}, where limited miss status handling registers (MSHR) resources are taken in the wrong order, and therefore must be reclaimed by instructions with earlier timestamps. 
	
	All three are uncommon; programs that transmit data ``backwards in time'' are unusual. Still, there are exceptions. Soplex often has \textit{timeleaps}, or load hits in the MSHRs, which are waiting for memory accesses to complete, where the original instruction that caused the MSHR to be allocated is logically after the new access to the cache in timestamp order, causing a replay of the load to ensure safety. This may be limited by more permissive \orderh\ mechanisms than the \lorder\ implemented in our prototype. Other workloads, such as mcf, libquantum, and omnetpp, experience \textit{leapfrogging}, caused by a combination of running out of MSHR resources, and MSHR resources being allocated in an order where later timestamped instructions take resources that would otherwise be allocated to those with an earlier timestamp. Still, the overheads could be reduced by providing more MSHRs in the cache.

	\subsection{Sizing Sensitivity}
	
\begin{figure*}[t]
		\centering
		\includegraphics[width=2\columnwidth]{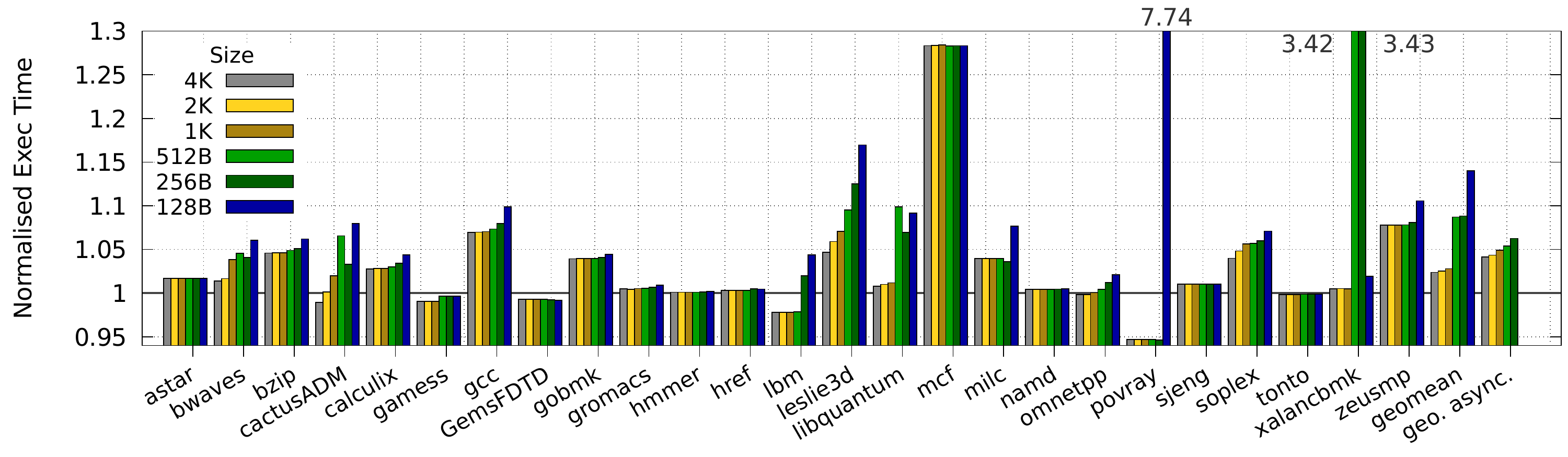}
		\caption{Slowdown of \name\ with different-sized instruction and data minions (2K default, each 2-way assoc.). Asynchronous reload of lost cache lines (\textit{geo. async.}, \response{where we only show the geomean for brevity}) removes the spikes in performance seen with small \name s, but is inadvisable for systems with ample resources.}
		\label{fig:size}
	\end{figure*}

	\Cref{fig:size} shows the performance of \name\ with various sizes of \name, where 2KiB is the default. We see that 4KiB is negligibly faster and 1KiB is negligibly slower. For some workloads, performance is relatively consistent regardless of \name\ size; most of their working set fits in the L1, and so use of the \name\ is rare in general. Others (e.g. leslie3d) feature a steady slowdown as the lack of space causes progressively more cache lines to be thrown away before reaching the non-speculative caches.
	
	Still, from 512B (8 cache lines) or fewer, spikes start to appear: in particular, povray at 128B, and Xalancbmk at 512B and 256B (but not 128B). In these cases, the \name\ is consistently not big enough, and so cache lines leave the \name\ before commit. This means they are repeatedly brought in from main memory. This can be solved by asynchronously reloading cache lines that were brought in to the \name\ by a speculative instruction, but are missing from the L1 cache and \name\ at commit time. This eliminates the spikes, and thus allows reasonable performance even with very small (128B, or 2 cache line) \name s. For more moderate sizes, this lowers performance, as it causes reload of some values that are never used again, especially in GemsFDTD and to a lesser extent in zeusmp, milc and lbm, so we turn it off by default.
	
	\subsection{Power Analysis}

	\name\ involves accessing the L1 cache and \name\ in parallel, unlike MuonTrap~\cite{Ainsworth_2020} where the speculative-hiding cache is an L0 filter cache. We have shown in \cref{sec:comparison} that this gives performance improvements when the speculative cache is frequently cleared, as is necessary for a comprehensive protection mechanism, but clearly there is an associated energy cost from the parallel accesses. Still, the timestamps that allow \name\ to safely use set associativity for its secret structures give significant efficiency compared with techniques that rely on full associativity or overprovisioning~\cite{yan2018invisispec,DBLP:journals/corr/abs-1806-05179} to hide storage contention.
	
	At 22nm, CACTI~\cite{muralimanohar2009cacti} places the static power consumption at $0.47mW$ per 2KiB \name, versus $12.8mW$ for the 64KiB L1 data cache, and read energy at $1.5pJ$ versus $8.6pJ$ for the L1 data cache, due to the latter's larger size. The new memory accesses from \name\ constitute of a read access in the \name\ for every read in the L1, a write for every line brought into the \name, and a read for every cache line brought into the \name\ and then committed into the L1. While some of this could be reduced by a form of way prediction to only access one of the two structures, provided this maintained the \order\ guarantees, the overhead is already small, with 3$\mu W$ maximum dynamic power draw for data, and 1$\mu W$ for instructions, across SPEC CPU2006. Compared with around $1W$ per core for recent Arm systems~\cite{PapadimitriouAdaptiveScaling}, the overheads are negligible. 
	
	
	\section{Related Work}
	
	Here we discuss existing (micro)architectural mitigations designed to prevent Spectre attacks. In turn, we consider how the concept of \order\ relates to their operation.
	
	\subsection{Speculation Hiding}
	
	Speculation-hiding techniques are intended to allow speculation to occur, but hide any effects from being observed by an attacker. There is varying compliance with \order, and thus vulnerability to contention attacks~\cite{behnia2020speculative,fustos2020spectrerewind}.
	
	InvisiSpec~\cite{yan2018invisispec} appears to offer \lorder\ (\cref{def:lorder}) for loads only: ``\textit{InvisiSpec reuses only state allocated by prior (in program order) instructions, to avoid creating new side channels.}'' To avoid incorrect eviction of data from its structures, it allocates them in program order. This requires overprovisioning, with full cache-line copies per each word-sized load-queue entry.  SafeSpec~\cite{DBLP:journals/corr/abs-1806-05179} appears to offer \order\  (\cref{def:order}): ``\textit{Speculative instructions in the same execution branch as the load that fetched a shadow cache line... can use the value from the shadow structure.}" It worst-case overprovisions its shadow structures to avoid contention from concurrent execution. Unlike InvisiSpec and \name, coherence channels~\cite{SpectrePrime} are not protected, and only \name{} handles MSHR and within-core~\cite{behnia2020speculative,fustos2020spectrerewind} contention. 
	\lorder{} directly translates into TimeGuarding in GhostMinion. This allows multiple levels of untrusted speculation to coexist within a set-associative cache structure, without leaking data via presence or absence. It gives more comprehensive enforcement, and without the overprovisioning of previous work~\cite{yan2018invisispec,DBLP:journals/corr/abs-1806-05179}.
	
	
	Data-oblivious execution~\cite{DataOblivious} makes all speculative instructions use a predicted amount of time on resources such as cache (by level prediction) and FPU (estimated cycle length). Under \order, many side channels are permitted without obliviousness: 
	earlier instructions in program order always allow \orderh\ transmission to later ones. 
	
	DAWG~\cite{kirianskydawg} dynamically partitions cache ways between distinct domains, such as groups of processes; MuonTrap~\cite{Ainsworth_2020} prevents the observation of speculative state, brought in to a filter cache, from affecting other processes, by flushing limited regions of speculation on context switch, and via coherence protocol changes. Neither enforces \order\ even between processes: though a cross-process attacker may not be able to directly observe speculative state changes, the timing of context switches and inter-process communication may still be altered by the effects of misspeculated execution. GhostMinion's TimeGuarding provides a direct method to augment these caches to support hiding concurrent execution, by preventing information flow backwards-in-time.
	
	Rollback techniques, such as CleanupSpec~\cite{Saileshwar:2019:CUA:3352460.3358314} and ReversiSpec~\cite{wu2020reversispec}, permit speculative execution to change the cache system, and roll it back under misspeculation. These violate \order\ on two counts. The time to roll back is dependent on the amount of state, which affects restarting of execution. As exploited in SpectreRewind~\cite{fustos2020spectrerewind} and Speculative Interference~\cite{behnia2020speculative}, correct execution, earlier in program order but concurrent with misspeculated execution, can see the effects of misspeculation. The latter property affects any techniques that only clear effects of misspeculation once it is discovered, such as MuonTrap-Flush~\cite{Ainsworth_2020}, Ghost Loads~\cite{10.1145/3310273.3321558}, and Gonzalez et al.'s RISC-V BOOM mitigation~\cite{gonzalez2019replicating}.
	
	
	
	
	Spectector~\cite{9152757} is a software analysis technique, based on the concept of \textit{speculative non-interference} of state leakage. DOLMA~\cite{loughlindolma} defines a similar property in \textit{Transient Non-Observability}. He et al.~\cite{he2020new} present the concept of attack graphs, which model access-control authorization delays on secret data in order to generate new vulnerabilities and evaluate defences against them. \order\ is a permissive constraint that can be used to define hardware that comprehensively obeys and defends against these properties.
	
	\subsection{Speculation Restricting}
	
	Alternatively, techniques can restrict speculative behaviour from occurring at all, rather than hiding its effects. Examples of such techniques include Speculative Taint Tracking~\cite{STT-micro}, NDA~\cite{NDA-micro}, Selective Delay and Value Prediction~\cite{SDVP}, Conditional Speculation~\cite{CondSpec}, and SpecShield~\cite{8714070}. Restricting unsafe speculation is sufficient to achieve \order, though it is not necessary. Still, restricting it entirely has fewer implications on complexity, state-hiding and coherence.
	To limit performance impact, such techniques violate \order\ in a selective way, for speculative instructions that can be demonstrated unlikely to leak secrets, for example not taking input from speculatively loaded data~\cite{STT-micro}.
	
	The fence instructions~\cite{intelnfo,taram2019context,armnonspec} built in to current microarchitectures can be viewed as a special case: these limit reordering of instructions around each side of the fence. Still, these could be implemented, without changing their guarantees, via \order\ around the barrier instead. 
	
	\subsection{Other Techniques}
	Other more ad-hoc speculation-restriction techniques are currently in use, such as  Retpoline~\cite{retpoline}, which redirects branch target flow, though these are harder to give precise \order\ analysis of. Isolation and randomisation of structures, such as over the branch target buffer~\cite{armv85aupdates} and branch predictor~\cite{10.1145/3404189}, restrict the ability of a cross-domain attacker from training a victim. Though they do not prevent attacks where the victim trains itself, and are thus more suited to protection against Variant 2~\cite{Kocher2018spectre} than other attacks, they provide strong probabilistic guarantees rather than full \order\ constraints.
	Kaiser~\cite{kaiser} is another example of isolation, where Meltdown~\cite{Lipp2018meltdown} is prevented by separating out the kernel and userspace's page tables, and thus gives protection outside of \orderh\ guarantees.

	\section{Conclusion}
	
	We have presented \order ing, a permissive constraint system designed to prevent speculative execution attacks, by comprehensively preventing the timing effects of mis-speculated instructions from leaking to instructions that do commit, either executed concurrently or in future. It does this while still allowing widespread forwarding of speculative information, by noting that, within a thread, the fates of many speculative instructions are tied together, and that processors generate instructions that become progressively more speculative, and so instructions earlier in program order can always produce side channels for those later.
	
	\name\ demonstrates the techniques needed to implement a full microarchitecture without heavy speculation restriction, while remaining safe via \orderh\ guarantees. By combining \order\ and \name\ with prediction for shared speculative resources between threads~\cite{DataOblivious}, and notions of Speculative Taint~\cite{STT-micro} to further permissify constraints, we believe systems will finally be able to rigorously eliminate transient-execution attacks at acceptable costs, ending this cat-and-mouse game for good.
	
\section*{Acknowledgements}
		\noindent This work was supported by the  Engineering and Physical Sciences Research Council (EPSRC), through grant reference EP/V038699/1.
		Additional data related to this publication is available at \url{https://github.com/SamAinsworth/reproduce-ghostminion-paper} or \url{https://doi.org/10.5281/zenodo.5222208}.

\balance
\small
\bibliographystyle{abbrv}

\end{document}